%% file: paper.tex
\documentclass[runningheads]{llncs}        
\usepackage{amssymb}
\usepackage{amsmath}

\usepackage{amsthm}
\usepackage{bussproofs}
\usepackage{stmaryrd}

\usepackage{dsfont}

\usepackage{bbm}
\usepackage{graphicx}
\usepackage{listings}
\usepackage{bussproofs}
\usepackage{tikz}
\usepackage{floatflt} 

\usepackage{ifthen}
\usepackage{xspace}
\usepackage{pdfpages}
\usepackage{todonotes}
\usepackage[inline]{enumitem}

\usepackage{floatflt} 
\usepackage{arydshln}
\usepackage{placeins}

\usepackage{pgf-umlsd}
\usetikzlibrary{arrows,shapes,trees,positioning,decorations,shapes}
\usetikzlibrary{decorations.markings,automata}

\input{abs}

\input{definitions}

\title{Deductive~Verification of Programs with Underspecified~Semantics by Model~Extraction}
\titlerunning{Deductive~Verification of Underspecified~Semantics}
\author{Eduard Kamburjan\orcidID{0000-0002-0996-2543}\inst{1,2} \and Nathan Wasser\inst{2}}

\institute{
   Department of Informatics, University of Oslo, Norway\\
  \email{eduard@ifi.uio.no}
  \and
   Department of Computer Science, Technische Universit{\"a}t Darmstadt, Germany\\
  \email{wasser@cs.tu-darmstadt.de}
}

\begin{document}
\maketitle

\begin{abstract}
We present a novel and well automatable approach to formal verification of programs with underspecified semantics, i.e., a language semantics that leaves open the order of certain evaluations.
First, we reduce this problem to non-determinism of distributed systems, automatically extracting a distributed Active Object model from underspecified, sequential C code.
This translation process provides a fully formal semantics for the considered C subset.
In the extracted model every non-deterministic choice corresponds to one possible evaluation order.
This step also automatically translates specifications in the ANSI/ISO C Specification Language (ACSL) into method contracts and object invariants for Active Objects. 
We then perform verification on the specified Active Objects model.
For this we have implemented a theorem prover \texttt{Crowbar} based on the Behavioral Program Logic (BPL), which verifies the extracted model with respect to the translated specification and ensures the original property of the C code for all possible evaluation orders. 
By using model extraction, we can use standard tools, without designing a new complex program logic to deal with underspecification.
The case study used is highly underspecified and cannot be verified with existing tools for C.
\end{abstract} 

\section{Introduction}\label{sec:intro}
\input{intro}

\section{Overview over Workflow}\label{sec:example}
\input{example}
\section{Abstract Behavioral Specification Language (ABS)}\label{sec:abs}
In this section we give the preliminaries for our work: the ABS language and cooperative contracts.
For space reasons, we refrain from introducing the full formalisms and refer to~\cite{Kamburjan19} for a full definition of the underlying program logic and to~\cite{soa} for a definition of the used ABS semantics and cooperative contracts. We stress, however, that the approach is fully formal.

\input{lang}
\subsection{Cooperative Method Contracts}\label{sec:bpl}
\input{bpl}
\section{Extraction of Annotated Model}\label{sec:c2abs}
\input{cabs}
\section{\texttt{Crowbar}\label{sec:impl}}
\input{crowbar}
\section{Case Study}\label{sec:case}
\input{fib}

\section{Conclusion}\label{sec:conclusion}
\input{conclusion}

\subsubsection*{Acknowledgments.}
This work was partially funded by the Hessian LOEWE initiative within the Software-Factory 4.0 project.

\bibliographystyle{acm}
\bibliography{ref}

\clearpage
\appendix

\paragraph{Proof of Lemma~\ref{lem2}.}
\begin{proof}
First, we observe that all extracted main block only create objects and call the \abs{call} methods on them.
For the following methods, \texttt{Crowbar} cannot deduce structural deadlock freedom.
\begin{enumerate}
\item \abs{C_one_to_fib.Int call_pred_or_id_fut_0(Fut<Int> fut_arg1)}
\item \abs{C_one_or_two.Int op_plus_fut_fut(Fut<Int> fut_arg1,Fut<Int> fut_arg2)}
\item \abs{C_pred_or_id.Int op_minus_val_fut(Int arg1,Fut<Int> fut_arg2)}
\item \abs{C_pred_or_id.Int op_plus_fut_fut(Fut<Int> fut_arg1,Fut<Int> fut_arg2)}
\item \abs{C_one_to_fib.Int op_plus_fut_fut(Fut<Int> fut_arg1,Fut<Int> fut_arg2)}
\item \abs{C_id_set_x.Int call(Int val)}
\item \abs{C_one_or_two.Int call()}
\item \abs{C_pred_or_id.Int call(Int val)}
\item \abs{C_one_to_fib.Int call(Int n)}
\end{enumerate}

The first 5 methods depends on futures as parameters. All passed futures are from structurally deadlock-free methods, \emph{because the main block does not call them}. 
Thus, all these methods is deadlock-free (i.e., not part of any deadlock).
The last 4 methods depend only on the first 5 methods and other structurally deadlock-free methods. Thus, all these methods are also deadlock-free.\qedhere
\end{proof}

\section{Formal Semantics of Contracts}\label{appendix:defs}
\input{appendix}
\subsection{Example of Local Traces}\label{appendix:examples}
\input{appendix_ex}
\clearpage
\section{Case Study: Extracted Annotated Model}\label{appendix:fib-abs}
\input{fib-abs}
\end{document}

%% file: abs.tex
\DeclareOldFontCommand{\rm}{\normalfont\rmfamily}{\mathrm}
\DeclareOldFontCommand{\sf}{\normalfont\sffamily}{\mathsf}
\DeclareOldFontCommand{\tt}{\normalfont\ttfamily}{\mathtt}
\DeclareOldFontCommand{\bf}{\normalfont\bfseries}{\mathbf}
\DeclareOldFontCommand{\it}{\normalfont\itshape}{\mathit}
\DeclareOldFontCommand{\sl}{\normalfont\slshape}{\@nomath\sl}
\DeclareOldFontCommand{\sc}{\normalfont\scshape}{\@nomath\sc}
\DeclareRobustCommand*\cal{\@fontswitch\relax\mathcal}
\DeclareRobustCommand*\mit{\@fontswitch\relax\mathnormal}

\colorlet{keywordcolor}{blue!50!black}
\colorlet{commentcolor}{green!60!black}
\colorlet{typecolor}{violet}
\newcommand{\sourcefont}{\ttfamily\small}
\newcommand{\commentfont}{\slshape\rmfamily\color{commentcolor}}

\lstdefinelanguage{ABS}{
        keywords={od,do,assert,this,new,type,def,case,of,local,class,interface,
        extends,implements,if,then,else,await,get,Fut,return,skip,while,module,
        import,export,from,to,suspend,delta,adds,modifies,removes,original,productline,
        features,core,corefeatures,optionalfeatures,after,when,product,hasAttribute,
        hasMethod,hasField,hasInterface,uses,root,extension,group,allof,oneof,require,
        stateupdate,object,main,objectupdate,classupdate,fi,
        exclude,original,ifin,ifout,opt,null,
        newgroup,thiscomp,in,joins,leaves,subtypeOf,acquire,except,as,component,Pre,Abs
        },
        keywordstyle=\color{keywordcolor}\bfseries\sffamily,
        morekeywords=[2]{Unit, Int, Bool, Rat, List, Set, Pair, Fut, Maybe, String, Triple, Either, Map},
        keywordstyle=[2]\color{typecolor},
        sensitive=true,
        comment=[l]{//},
        morecomment=[s]{/*}{*/},
        morestring=[b]"
}

\lstdefinelanguage[v9]{Java}[]{Java}{
        morekeywords={module,requires,provides,uses,with,to,exports}
}
\lstdefinelanguage[ContextJ]{Java}[]{Java}{
        morekeywords={layer,with,without,proceed,before,after}
}
\lstdefinelanguage[FOP]{Java}[]{Java}{
        morekeywords={refines,original,Super}
}
\lstdefinelanguage[JastAdd]{Java}[]{Java}{
        morekeywords={aspect,syn,inh,lazy}
}

\lstdefinestyle{code}{
        basicstyle=\sourcefont\upshape,
        keywordstyle=\color{keywordcolor}\bfseries\sffamily,
        commentstyle=\commentfont,
        columns=fullflexible,
        mathescape=true,
        escapechar={\#},
        keepspaces=true,
        showstringspaces=false,
        aboveskip=8pt, 
        numbers=left,
        stepnumber=1, 
        numberstyle=\ttfamily\scriptsize\color{gray},
        numbersep=4pt,
        xleftmargin=1.5em,
        xrightmargin=1.5em,
        framexleftmargin=1.2em,
        framexrightmargin=1em,
        framextopmargin=0.5ex,
        breaklines=true,
        breakindent=3pt,
}

\lstdefinestyle{abs}{
        style=code,
        language=ABS,
}
\lstdefinestyle{C}{
        style=code,
        language=C,
}
\lstdefinestyle{java}{
        style=code,
            language=Java
}
\lstdefinestyle{java9}{
        style=code,
            language=[v9]Java
}
\lstdefinestyle{aspectj}{
        style=code,
        language=[AspectJ]Java
}
\lstdefinestyle{jastadd}{
        style=code,
        language=[JastAdd]Java
}
\lstdefinestyle{contextj}{
        style=code,
        language=[ContextJ]Java
}
\lstdefinestyle{FOP}{
        style=code,
        language=[FOP]Java
}
\lstdefinestyle{scala}{
        style=code,
        language=Scala,
        morekeywords={self}
}

\newcommand{\code}[2][]{\lstinline[style=code,basicstyle=\ttfamily\upshape,#1]{#2}}

\newcommand{\abs}[2][]{\code[style=abs,#1]{#2}}

\lstnewenvironment{srccode}[1][]{
        \minipage{\linewidth}
        \lstset{style=code,
        backgroundcolor=\color{white},
        rulecolor=\color{gray!50},
        frame=tblr,
        captionpos=b,
        #1}
}{\endminipage}

\lstnewenvironment{abscode}[1][]{
        \minipage{1\linewidth}
        \lstset{style=abs,
        backgroundcolor=\color{white},
        rulecolor=\color{gray!50},
        frame=tblr,
        captionpos=b,
        #1}
}{\endminipage}

\lstnewenvironment{ccode}[1][]{
        \minipage{1\linewidth}
        \lstset{style=C,
        backgroundcolor=\color{white},
        rulecolor=\color{gray!50},
        frame=tblr,
        captionpos=b,
        #1}
}{\endminipage}

\lstnewenvironment{javacode}[1][]{
        \minipage{\linewidth}
        \lstset{style=java,
        backgroundcolor=\color{gray!5},
        rulecolor=\color{gray!50},
        frame=tblr,
        captionpos=b,
        #1}
}{\endminipage}

\lstnewenvironment{jastaddcode}[1][]{
        \minipage{\linewidth}
        \lstset{style=jastadd,
        backgroundcolor=\color{gray!5},
        rulecolor=\color{gray!50},
        frame=tblr,
        captionpos=b,
        #1}
}{\endminipage}

%% file: definitions.tex
\newcommand{\EK}[1]{#1}

\newcommand{\NW}[1]{#1}

\newcommand{\HIDE}[1]{}

\makeatletter
\newsavebox{\@brx}
\newcommand{\llangle}[1][]{\savebox{\@brx}{\(\m@th{#1\langle}\)}%
  \mathopen{\copy\@brx\kern-0.5\wd\@brx\usebox{\@brx}}}
  \newcommand{\rrangle}[1][]{\savebox{\@brx}{\(\m@th{#1\rangle}\)}%
    \mathclose{\copy\@brx\kern-0.5\wd\@brx\usebox{\@brx}}}
    \makeatother

\let\temp\phi
\let\phi\varphi
\let\varphi\temp

\makeatletter
\newcommand{\xRightarrow}[2][]{\ext@arrow 0359\Rightarrowfill@{#1}{#2}}

\DeclareMathOperator*{\halfsim}{\Vdash}

\newcommand{\xabs}[1]{\text{\abs{#1}}}

\newcommand{\sep}{  \ | \ }
\newcommand{\many}[1]{\overrightarrow{#1}}

\newcommand{\sequence}[1]{\ensuremath{ \left\langle #1 \right\rangle \xspace}}
\newcommand{\bigsequence}[1]{\ensuremath{ \big\langle #1 \big\rangle \xspace}}

\newcommand{\methodname}{\ensuremath{\synt{m}}\xspace}

\newcommand{\invocev}{\ensuremath{\mathsf{invEv}}\xspace}
\newcommand{\invocrev}{\ensuremath{\mathsf{invREv}}\xspace}
\newcommand{\resolvev}{\ensuremath{\mathsf{futEv}}\xspace}
\newcommand{\resolvrev}{\ensuremath{\mathsf{futREv}}\xspace}
\newcommand{\noev}{\ensuremath{\mathsf{noEv}}\xspace}
\newcommand{\awaitev}{\ensuremath{\mathsf{condEv}}\xspace}
\newcommand{\awaitrev}{\ensuremath{\mathsf{condREv}}\xspace}
\newcommand{\suspev}{\ensuremath{\mathsf{suspEv}}\xspace}
\newcommand{\susprev}{\ensuremath{\mathsf{suspREv}}\xspace}

\HIDE{

}
\def\fCenter{\ \Rightarrow\ }

\newcommand{\synt}[1]{\ensuremath{\xabs{#1}}}

\newcommand{\type}{\ensuremath{\mathbbm{T}}\xspace}

\newcommand{\trace}{\ensuremath{\theta}\xspace}

\newcommand{\marker}{\ensuremath{\diamond}\xspace}

\newcommand{\eval}[2]{\ensuremath{ \left\llbracket #1 \right\rrbracket_{#2} \xspace}}

\newcommand{\inv}{\ensuremath{\mathsf{inv}}\xspace}

%% file: intro.tex
Verification of programs relies on the availability of a formal, or at least a formalizable, semantics of the used programming language. 
\EK{If verification is available, it can be used to automatize engineering tasks such as parallelization of legacy software~\cite{HahnleTMNS020}.}

\begin{figure}
\noindent\begin{minipage}{\columnwidth}\begin{ccode}
int x;
int id_set_x(int val){
  x=1;
  return val;
}
int main(void){
  x=0;
  return x + id_set_x(1);
}
\end{ccode}
\end{minipage}
\caption{\label{fig:cpure} Addition with side-effect.}
\end{figure}
In this work we consider the semantics of the C language, which in addition to fully specified behavior contains \emph{undefined}, \emph{unspecified} and \emph{implementation defined} behavior: these semantics describe not exactly what should happen, but leave crucial decisions to the implementing compiler and/or the runtime environment.
Our focus here is on the unspecified evaluation order within the C standard, which we refer to as \emph{underspecified}.
Importantly, the semantics for underspecified behavior is not \emph{undefined}, as the semantics limits the possible choices.
This is not merely a fringe case, but is observable already in natural and small programs. Consider the C program in Fig.~\ref{fig:cpure}.
The C99 standard~\cite{ISO:C99} does not specify the order of evaluation of the subexpressions in the addition.%
\footnote{%
This unspecified evaluation order is also prevalent in other C standards.}
Indeed, the two main compilers for C return different values: \texttt{gcc} 7.4.0 returns 2 (evaluating the second summand first), \texttt{clang} 6.0.0 returns 1 (evaluating the first summand first).
The reason is that \texttt{gcc} uses a stack-based translation of expressions, while \texttt{clang} uses a queue-based one.

Verification of underspecified C code is still an open problem and merely \emph{fixing} the choice is not enough for verification:
As the semantics is underspecified, compilers are not required to be consistent in their choice \emph{even during the run of a single program} and optimizations are not obligated to preserve the choice of the compiler.

This effect is further amplified from a software engineering perspective, when program equivalence becomes a problem: 
For one, changing or updating the compiler, or indeed barely changing its parameters may result in different program behavior.
For another, reengineering legacy software, a critical activity to, e.g., make use of parallelization~\cite{HahnleTMNS020} cannot rely on analyses proving functional equivalence, if these analyses are not considering underspecified semantics.
Before attempting to prove program equivalence, one must be able to reason about functional behavior of programs in a language with underspecified semantics.

This work presents an approach to \emph{automatically} verify functional behavior of C programs with underspecified semantics,
which is based on reducing \emph{underspecification} to \emph{non-determinism} in a fully specified language:
We are able to verify functional properties of C programs without undefined behavior with respect to every possible standard-compliant semantics. 
In this work we build upon the model-extraction approach by Wasser et al.~\cite{WasserHH19} for a subset of the C language
and give an \emph{implemented} system that verifies the functional behavior of the extracted model.
The extracted model gives a \emph{fully formal} and analyzable semantics for C in terms of an Active Object framework.

We translate C code into an \emph{Active Objects} language~\cite{boer} and regard sequential C programs as parallel programs, 
in which the non-determinism arises from parallelism and not from underspecified semantics.
Conceptually, this is a rare case where a problem of \emph{sequential} programs is transformed to a problem of \emph{parallel} programs, because the support for analysis of parallel systems is better than the support for reasoning about underspecified semantics.

For Active Objects there are program logics~\cite{Kamburjan19} that enable modular reasoning and we are able to employ method contracts for asynchronous calls~\cite{soa}. 
The expected behavior under all possible semantics is annotated with ACSL~\cite{acsl} and automatically translated into cooperative contracts and object invariants of Active Objects.
Using this approach we give a case study to verify that a highly underspecified recursive function that computes the $n$th Fibonacci number in one semantics
returns a value between $1$ and the $n$th Fibonacci number in every standard-adhering semantics. 

\paragraph*{Contributions}
Our contributions are 
(1) an implemented\footnote{Tool and examples available at \url{formbar.raillab.de/c2abs2bpl}.} approach to \emph{automatically} verify functional behavior of C programs with underspecified semantics,
and a deductive verification case study of underspecified C code which is 
(2) the biggest verification case study of such code that cannot be handled by existing approaches (see next section)
(3) the biggest deductive verification case study for Active Objects (in lines of code) to date.
The case study can be proven \emph{fully automatically}.
Furthermore, we present (4) the \emph{first} implementation of BPL in the \texttt{Crowbar} prover.

\paragraph*{State-of-the-Art and Related Work}
Underspecified (and to a lesser degree undefined) semantics are a rarely approached challenge for deductive verification tools. Here, we review the tools that consider these kinds of semantics.

%
Frama-C~\cite{framac} can find (some) \emph{undefined} behavior related to read-write or write-write accesses \emph{between} sequence points. However, it does not recognize \emph{unspecified} behavior when these accesses occur \emph{indeterminately sequenced} as in our examples here, instead only examining a single fixed evaluation order\footnote{E.g., value analysis in Frama-C claims that the program in Fig.~\ref{fig:cpure} can only return 2.}~\cite[p.40]{framamanual}.
%
Further, while most of ACSL is utilized in Frama-C, this does not include global invariants, which we are able to handle.
Additionally, new tools must be built specifically for the C intermediate representation only used within Frama-C, while our approach can profit from all tools available for ABS, which has included so far model checking, simulation, deadlock analysis and deductive verification.
%
%
RV-Match~\cite{rv-match-url}---based on C semantics formalized~\cite{k-framework-c-semantics-url,k-framework-c-semantics} in the K framework~\cite{k-framework-url,k-framework}---is able to find (some) \emph{undefined} and \emph{implementation defined} behavior in C programs, but like Frama-C chooses only a single evaluation order when faced with \emph{underspecified} behavior.
This in turn prevents both from finding undesired behavior that is only obvious when a different evaluation order is chosen.
While our approach currently works only with an admittedly smaller subset of C containing underspecification than that allowed in RV-Match and Frama-C, it faithfully considers all possible evaluation paths allowed by the standard.
%
%
Cerberus~\cite{cerberus-url,cerberus} is an analysis tool for undefined and underspecified behavior; however, it cannot utilize any specifications and its treatment of unspecified evaluation order of \emph{side effects} does not match the C standard, as demonstrated in~\cite{WasserHH19}.
%
%
The separation logic system of Frumin et al.~\cite{FruminGK19}, based on small-step semantics in Coq~\cite{Krebbers14} correctly treats underspecification.
They give a formal system to verify a program in their toy language $\lambda$MC and check effects of underspecified behavior with a modified separation logic.
In contrast to the subset of C we consider, $\lambda$MC is emphatically \emph{not} a subset of C and is described as merely a C-style language\footnote{Even this is debatable, but underspecified C-style behavior is present.}.
Verification of \emph{any} C program therefore requires manual translation into an equivalent $\lambda$MC program and manual specification of the $\lambda$MC program in Coq.
Our model-extraction based approach is fully automated, can be used with standard program logics and analyses for Active Objects and does not rely on complex rule modifications to handle underspecified behavior. \NW{We stress that this automation includes the verification
, which needs not be performed by the user in an interactive prover such as Coq~\cite{Coq-url}.}

\NW{Holzmann and Smith~\cite{HolzmannS02} attempt to reuse the SPIN model checker by extracting Promela code from a C program.
However, their approach requires \emph{manual} translation/adjustment (flattening) of the underspecified parts. 
Furthermore, Promela/SPIN only support model checking and cannot be applied to unbounded inputs.}

Concerning semantics, several formalizations~\cite{ellison2012,norrish1998,papaspyrou2001} of the C semantics deal with underspecified evaluation order without giving a reasoning system.

To conclude the overview of the state-of-the-art, there is no satisfying approach to verify underspecified C code and the partial approaches are not suited for \NW{automatization}.

\paragraph*{Structure.}
In Sec.~\ref{sec:example} we investigate the program in Fig.~\ref{fig:cpure} in detail.
In Sec.~\ref{sec:abs} we give preliminaries: the basics of ABS~\cite{abs}, the Active Object language used, and of cooperative method contracts.
In Sec.~\ref{sec:c2abs} we describe the model-extraction and in Sec.~\ref{sec:impl} the implementation of \texttt{Crowbar}, which are then used in Sec.~\ref{sec:case} to verify the Fibonacci case study. We conclude in Section~\ref{sec:conclusion}.

%% file: example.tex
Before we introduce the needed formalisms, we give a short overview over our approach using the code in Fig.~\ref{fig:example1}, which adds ACSL specifications to the previous example.
The strong global invariant specifies a condition that has to hold at every point during execution, while the requires/ensures clauses are standard pre/postconditions.
\begin{figure}[b]
\begin{ccode}[basicstyle=\fontsize{9}{10}\ttfamily,keywordstyle=\color{keywordcolor}\bfseries\sffamily]
int x; //@ strong global invariant x == 0 || x == 1;$\label{strong_global_invariant}$
int id_set_x(int val) 
/*@ requires val == 1; ensures \result == 1; @*/ {$\label{id_set_x_spec}$
  x=1; return val;
}
int main(void) 
/*@ ensures \result == 1 || \result == 2; @*/ {
  x=0; return x + id_set_x(1);
}
\end{ccode}
\caption{Specified addition with side-effect.}
\label{fig:example1}
\end{figure}

Specified C-code is translated into specified ABS-code. ABS is object-oriented and uses a concurrency model with the following features to model concurrency:
(1) An object cannot access the fields of another object. (2) Every method call is asynchronous (i.e., does not block the caller) and returns a future. A future can be used to synchronize on the called method and read its eventual return value. (3) Only one process is active per object and a process can only be interrupted when executing an \abs{await g} statement.
An \abs{await g} statement waits until all futures in the guard \abs{g} are resolved, i.e., their process has terminated.
As for specifications, ABS supports object invariants and pre/postconditions on methods. There are no global variables.

The code in Fig.~\ref{fig:example2} shows a (prettified) part of the translation of Fig.~\ref{fig:example1}.
The global variables are handled by a special (singleton) class \abs{Global}.
In \abs{Global}, each global variable is a field and the global invariant becomes the object invariant of this class.
Similarly, the global invariant is also added as pre/postcondition to the setter and getter method handling the fields.

\begin{figure}[bt]
\begin{abscode}[basicstyle=\fontsize{9}{10}\ttfamily,keywordstyle=\color{keywordcolor}\bfseries\sffamily]
[Spec:ObjInv(this.x == 0||this.x == 1)]
class Global implements Global {
 Int x = 0;
 [Spec:Ensures(result == 0||result == 1)]
 Int get_x() { return this.x; }
 [Spec:Requires(value == 0||value == 1)]
 Unit set_x(Int value) { this.x = value; }
}
class C_id_set_x(Global global) 
      implements I_id_set_x {
 [Spec: Requires( val == 1 )] 
 [Spec: Ensures( result == 1 )] 
 Int call(Int val){...}// executes id_set_x(val)
 ... }
class C_main(Global global) 
      implements I_main {
 [Spec:Ensures(result == 1||result == 2)]
 Int call() { // executes main()
  Fut<Unit> tmp_4 = 
    this!set_global_x_val(0); // sets x to 0
  await tmp_4?; // introduces sequence point ``;''$\label{await_after_setting_x}$
  Fut<Int> tmp_5 = 
    this!get_global_x(); // reads x
  Fut<Int> tmp_6 = 
    this!call_id_set_x_val_0(1);//calls id_set_x
  Fut<Int> tmp_7 = 
    this!op_plus_fut_fut(tmp_5, tmp_6);//add
  await tmp_7?; // introduces sequence point ``;''
  return tmp_7.get; // returns result of addition
 }
 [Spec: Ensures(valueOf(fut_arg1) + valueOf(fut_arg2) == result)]
 Int op_plus_fut_fut(Fut<Int> fut_arg1, 
                     Fut<Int> fut_arg2) {
  await fut_arg1? & fut_arg2?; $\label{line1}$
  Int arg1 = fut_arg1.get; 
  Int arg2 = fut_arg2.get;
  return ( arg1 + arg2 );
 }
 ... }
\end{abscode}
\caption{Partial translation of Fig.~\ref{fig:example1}.}
\label{fig:example2}
\end{figure}

Each C-function \abs{f} is translated into an ABS-class \abs{C_f} and an interface \abs{I_f} with a \abs{call} method that models its execution.
The function contract of \abs{id_set_x} becomes the method contract of \abs{I_id_set_x.call}.

We only show the translation of \abs{main} in detail.
Again, the function contract becomes the method contract of \abs{call}.
The other methods in the class \abs{C_main} model memory accesses to global variable \abs{x}, calling function \abs{id_set_x} and addition with the \abs{+} operator.

The \abs{call} method is a translation of the \abs{main} function. 
It first sets \abs{x} to 0 and than waits for this operation to finish --- the \abs{await} at line~\ref{await_after_setting_x} models synchronization at the sequence point \abs{;}.
The next three lines translate the addition operation and contain \emph{no} \abs{await}, because the C-expression contains no sequence point.
The two calls to model evaluation of the subexpressions are called in one order, but may be executed in a different one. 

The method \abs{op_plus_fut_fut} models evaluation of the addition expression. It takes two futures, i.e., two references to \emph{yet unfinished executions}.
It then synchronizes with both of them, i.e., it waits until \emph{both} are resolved (line~\ref{line1}) and then adds the corresponding return values.
It depends on the global scheduling which method is executed first and therefore whether the read triggered in \abs{C_main} or the write in \abs{C_id_set_x} takes place on \abs{Global} first.
Note that the specification of \abs{C_main} is also automatically derived from the ACSL specification.

The translated model can now be passed to the \texttt{Crowbar} verification system, which checks that the code adheres to its specification.
\FloatBarrier

%% file: lang.tex
ABS~\cite{abs} is an executable, object-oriented modeling languages based on Active Objects~\cite{boer}, designed to model and analyze distributed systems.
It has been applied to model a wide range of concurrent software systems, such as cloud-based services~\cite{fred}, YARN~\cite{yarn} or memory systems~\cite{wao}.

\subsection{Overview.} ABS syntax is largely based on Java and we refrain from describing the full language here.
Instead, we introduce ABS in an example-driven way to demonstrate its concurrency model and formal semantics.
The main features of the concurrency model can be summarized with the points below:
\begin{description}
\item[Strong Encapsulation.]
Every object is strongly encapsulated at runtime, such that no other object can access its fields, not even objects of the same class.
\item[Asynchronous Calls with Futures.] 
The ABS language combines actors~\cite{actor} with futures~\cite{future}. 
Each method call is asynchronous and generates a future. Futures can be passed around and are used to synchronize on the process generated by the call.
Once the called process terminates, its future is \emph{resolved} and the return value can be retrieved. We say that the process \emph{computes} its future.
\item[Cooperative Scheduling.] 
At every point in time, at most one process is active in an object. Active Objects are preemption-free: A running process cannot be interrupted unless it \emph{explicitly} releases control over the object.
This is done either by termination with a \abs{return} statement or with an \abs{await g} statement that waits until guard \abs{g} holds.
A guard polls a set of futures 
and holds iff all futures in it are resolved.
\end{description}

These features ensure that a process has exclusive control over the heap memory of its object between syntactically marked statements. 
This vastly simplifies deductive verification, as between such statements techniques from sequential program verification carry over directly.

\begin{example}\label{ex:lang}
As the extracted models from C code are rather unintuitive, we demonstrate the concurrency model of ABS with a more natural program.

Fig.~\ref{fig:absintro} gives an ABS model with two objects that folds some binary operation over three numbers: one object that performs the operation and a second object that performs the folding.
Interface \abs{Fold} defines an interface for the fold.
Lines~\ref{line1-2} and \ref{line2} give the specification, which we discuss in more detail in section~\ref{sec:bpl}. 
Here, we specify that the input values must be positive (\abs{Requires}) and that the result is positive (\abs{Ensures}).
Interface \abs{Comp} specifies a single method, which performs some operation that also operates only on positive numbers.
Class \abs{FoldC} implements the folding and has a field \abs{comp} that points to a \abs{Comp} instance. We specify that the field is initialized with a non-null value (\abs{Requires}) 
and stays non-null (\abs{ObjInv}).
It has a field \abs{last} to store the intermediate result.
%
ABS uses a main block to initialize the system, which here creates one instance of each class, starts two \abs{fold}-processes and synchronizes on both.
There is no await in the class -- the processes executing \abs{C.fold} do not overlap, so the value of \abs{last} cannot change before it is returned and it is safe to save the intermediate value in this field. 
\end{example}

\begin{figure}[t!b]
\begin{abscode}[basicstyle=\fontsize{9}{10}\ttfamily,keywordstyle=\color{keywordcolor}\bfseries\sffamily]
interface Fold { 
  [Spec: Requires(a>0 && b>0 && c>0)]#\label{line1-2}#
  [Spec: Ensures(result>0)]#\label{line2}#
  Int fold(Int a, Int b, Int c);
}
interface Comp { 
  [Spec: Requires(a>0 && b>0)]
  [Spec: Ensures(result>0)]
  Int op(Int a, Int b);
}
class CompC implements Comp { ... }
[Spec: Requires(comp != null)]
[Spec: ObjInv(comp != null)] 
class FoldC(Comp comp, Int last) 
      implements Fold{	
  Int fold(Int a, Int b, Int c){
    Fut<Int> f = comp!op(a, b); last = f.get;
    f = comp!op(last, c); last = f.get;
    return last;
  }
}
{ Comp a = new CompC(); 
  Fold c = new FoldC(a,0); 
  Fut<Int> f1 = c!fold(1,2,5); 
  Fut<Int> f2 = c!fold(1,2,4);
  await f1? & f2?; }
\end{abscode}
\caption{Simple ABS Model, slightly beautified.}
\label{fig:absintro}
\end{figure}

\subsection{Semantics.}
ABS uses a \emph{Locally Abstract, Globally Concrete} (LAGC) trace semantics~\cite{lagc} as its formal foundation.
LAGC semantics allow one to give simpler soundness arguments for program logics, than SOS semantics.
An LAGC semantics consists of two layers: a denotational semantics for methods that maps each method to a set of local traces and an operational semantics for objects and the whole system that combines local traces to global traces.
For space reasons, we only present the main ideas here, for a full definition we refer to~\cite{soa}.

A local trace is an alternating sequence of symbolic states and events.
A state maps every field and variable to a symbolic value and an event is a marker for a visible communication.
A local trace is concrete if all its symbolic values are replaced by concrete values.
The semantics of a statement is a set of symbolic local traces, which describes all possible concrete local traces. 
However, the symbolic values allow us to reason about the method in isolation, while concrete traces can only be generated if the concrete context is known.

\begin{example}[Symbolic Trace]
Consider the statement \xabs{i = j+1; o!m(); return i}.
Its semantics is the following, slightly simplified, symbolic trace, where $\overline{\cdot}$ denotes symbolic values.

\noindent\scalebox{0.9}{
\begin{minipage}{\columnwidth}
\begin{align*}
\Big\langle 
&\big(\xabs{o}\mapsto\overline{o},\xabs{i} \mapsto \overline{i},\xabs{j} \mapsto \overline{j}\big),\noev,\big(\xabs{o}\mapsto\overline{o},\xabs{i} \mapsto \overline{j}+1,\xabs{j} \mapsto \overline{j}\big),\\
&\mathsf{callEv}(\xabs{this},\overline{o},\xabs{m}),\big(\xabs{o}\mapsto\overline{o},\xabs{i} \mapsto \overline{j}+1,\xabs{j} \mapsto \overline{j}\big),\mathsf{returnEv}(\overline{j}+1)
\Big\rangle
\end{align*}
\end{minipage}}

Here, \noev denotes a step without a visible effect, $\mathsf{callEv}$ a method call to a symbolic callee $\overline{o}$ and $\mathsf{returnEv}$ a termination with a given return value.
\end{example}

The global semantics generate concrete traces by gradually instantiating concrete values for symbolic values.
The advantage of LAGC semantics are, for one, simple soundness arguments for program logics, as it splits the local analysis of single methods does not need to reason about the SOS semantics~\cite{abs}. For another, it allows to express trace properties locally in a very natural way in a dynamic logic, as we show in Sec.~\ref{sec:bpll}.

%% file: bpl.tex
Here, we give the used fragment of the specification language for ABS: cooperative method contracts~\cite{soa} and object invariants for Active Objects~\cite{DinO15}.
We recap the Behavioral Program Logic~\cite{Kamburjan19} used to verify cooperative method contracts. 

Cooperative Method Contracts use two kinds of preconditions for methods: \emph{parameter preconditions}, which describe the expected parameters; and \emph{heap preconditions}, which additionally describe the class fields.
Splitting the precondition is necessary, because the parameters are controlled by the \emph{caller process} (and must be guaranteed by the caller), 
while the fields are controlled by the last active process in the \emph{callee object} (and must be guaranteed by this process).
There are also two postconditions: the heap postcondition
defines the final state upon termination of the method in terms of its fields and local variables plus a special program variable \abs{result} for the return value;
the parameter postcondition defines the return value in terms of the parameters. The parameter postcondition can be used upon reading from the future if the call parameters are known.
%

We also use object invariants, which must hold at every point a method loses or regains control over the object: at method start, termination and \abs{await} statements.
The initial state of classes is specified with \emph{creation conditions}. 

\paragraph{Specification.}
Method signatures in interfaces may be annotated with parameter preconditions \mbox{\abs{[Spec:Requires(e)]}} and postconditions \mbox{\abs{[Spec:Ensures(e)]}}, where \abs{e} is an expression of Boolean type.
Similarly, method implementations in classes may be annotated with heap pre- and postconditions.
A heap precondition that could be a parameter precondition is automatically transformed.
Classes may be annotated with object invariants \mbox{\abs{[Spec: ObjInv(e)]}} and creation conditions \mbox{\abs{[Spec: Requires(e)]}}.
Loops may be annotated with loop invariants \mbox{\abs{[Spec: WhileInv(e)]}}.
The specifications in Fig.~\ref{fig:absintro} are explained in Example~\ref{ex:lang}.

Full cooperative contracts have mechanisms to specify and verify \abs{await} statements with suspension contracts and \abs{get} statements with resolving contracts~\cite{soa}. 
Similarly, so called \emph{context sets}~\cite{soa} are used to specify and analyze the heap preconditions.
As neither heap preconditions nor suspension or resolving contracts are used by the extracted models, we refrain from introducing them in detail.


\subsection{BPL for Cooperative Method Contracts}
\label{sec:bpll}
\paragraph{Syntax and Semantics.}
BPL generalizes dynamic first-order logic and is parametric in the verified property.
A behavioral specification $\mathbb{T}$ is a pair $(\alpha_\type,\tau_\type)$, where $\alpha_\type$ is a map from $\tau_\type$ to formulas in a trace logic (not BPL).
\begin{definition}[Syntax of BPL]
Let $(\alpha_\type,\tau_\type)$ be a behavioral specification.
Let $p$ range over predicate symbols, $f$ over function symbols, \abs{v} over program variable, $x$ over logical variables, $S$ over sorts and $\tau$ over $\tau_\type$.
Updates $U$, formulas $\phi$ and terms $t$ of BPL are defined by the following syntax.
\begin{align*}
\phi &::= p(\many{t}) \sep \neg \phi \sep \phi \vee \phi \sep \exists x \in S.~\phi \sep \big[\xabs{s} \halfsim^{\alpha_\mathbb{T}} \tau\big] \sep \{U\}\phi\\
t &::= x \sep \xabs{v} \sep f(\many{t}) \sep \{U\}t \qquad U ::= \xabs{v} := t \sep U || U \sep \{U\}U
\end{align*}
\end{definition}
Intuitively, an update $\xabs{v} := t$ is an explicit substitution of $t$ for $\xabs{v}$. We use updates to delay state updates when symbolically executing a statement.
$U||U$ denotes accumulation of two updates and $\{U\}\cdot$ denotes an application of the substitution. 
The modality $\big[\xabs{s} \halfsim^{\alpha} \tau\big]$ expresses that every trace of \abs{s} is a model in a trace logic (defined below) for $\alpha_\mathbb{T}(\tau)$.
This is a generalization of postcondition reasoning where $\tau$ would be a formula and $\alpha$ a trivial translation into trace logic.
Using behavioral specifications allows us to design the calculus without explicitly modeling the trace.
Field access is modeled via a special program variable \abs{heap} and the usual theory of arrays with select and store functions~\cite{BeckertK016}.
Formally we give the semantics as follows.
\begin{definition}[Semantics of BPL]
The semantics of a behavioral modality is given below as a satisfiability relation $\sigma,I,\beta \models \phi$, where $\sigma$ is a state, $\beta$ a variable assignment and $I$ an interpretation of function and predicate symbols.
\[\sigma,I,\beta \models \big[\xabs{s} \halfsim^\alpha \tau\big] \text{ iff } \forall \trace \in \eval{\xabs{s}}{\sigma}.~\trace,I,\beta \models \alpha(\tau)\]
Where $\eval{\xabs{s}}{\sigma}$ is the set of concrete local traces that result from the instantiating the symbolic traces of $\xabs{s}$ according to the initial state $\sigma$.
The semantics $\eval{t}{\sigma,\beta}^I$ of a term is a domain element and the semantics $\eval{U}{\sigma,\beta}^I$ of an update is a map from states to states.
Both are standard~\cite{BeckertK016,Kamburjan19}, as is the semantics of first-order connectives. 
We stress that BPL is defined on concrete trace, where states are simple first-order logic models and events are tuples.
\end{definition}
We stress that $\alpha(\tau)$ maps to a different logic than BPL.
In this work, we use a combined behavioral specification for object invariants, post-conditions and contracts, which we define next.
\begin{definition}[Contracts in BPL]
The syntax of the behavioral specification is the set of behavioral contracts.
A behavioral contract for a method \xabs{C.m} is a tuple $\inv,M,\phi,\psi$, where 
\begin{itemize}
\item \inv is a modality-free BPL formula that contains only fields from \abs{C} and no program variables except \abs{heap},
\item $M$ is a map from method names to pairs of modality-free BPL formulas. The first element $M^\mathsf{pr}(\xabs{m'})$ is the parameter precondition and the second element $M^\mathsf{po}(\xabs{m'})$ the parameter postcondition.
\item $\phi$ is a modality-free BPL formula that contains only fields from \abs{C}, only program variables from \abs{m}, and the special variables \abs{heap} and \abs{result},
\item $\psi$ is a modality-free BPL formula that contains only fields from \abs{C} and no program variables except \abs{heap}.
\end{itemize}
\end{definition}
\noindent Formula \inv is the object invariant and $\phi$ the postcondition of the method. Both are given as user annotations.
Map $M$ models the contracts of called methods. Finally, $\psi$ is the \emph{statement} postcondition. The split between 
$\phi$ and $\psi$ is necessary to enable $\psi$ to prove the loop invariant at the end of the loop body.

The semantics $\alpha(I,M,\phi,\psi)$ is a first-order formula in a \emph{trace logic}. Contrary to BPL, the models of the trace logic are local traces. The events of the local traces are used to connect state and trace. E.g., the method precondition is expressed as the specification of the state after a \resolvev event. Due to limited space, we refer to the formal treatment of the semantics $\alpha$ to~\cite{KamburjanDHJ19}. 

Each method $\xabs{C.m}$ is verified against a proof obligation 
\[
\inv_\xabs{C.m} \wedge \mathsf{pr}_\xabs{C.m} \rightarrow \big[ \xabs{s}_\xabs{C.m} \halfsim^{\alpha} \inv_\xabs{C.m},M_\xabs{C.m},\phi_\xabs{C.m},\mathsf{true}\big]
\]
where $\mathsf{pr}_\xabs{C.m}$ is the conjunction of all preconditions in class and interfaces for this method, $I_\xabs{C.m}$ is the annotated object invariant for \abs{C},
$\phi_\xabs{C.m}$ the conjunction of all postconditions in class and interfaces for this method and $M_\xabs{C.m}$ maps every method name to a pair of its precondition and its postconditions annotated in the interfaces.
Additionally, there is a proof obligation to check that object precondition and initialization of the fields imply the object invariant.
\begin{example}
The proof obligation for \abs{FoldC.fold} from Fig.~\ref{fig:absintro} is as follows:
\begin{align*}
&\big(\xabs{comp} \not\doteq \xabs{null} \wedge \xabs{a} >  0 \wedge \xabs{b} >  0 \wedge\xabs{c} >  0 \big) \\
\rightarrow&\big[ \dots \halfsim^{\alpha} \xabs{comp} \not\doteq \xabs{null},M,\xabs{result} >  0,\mathsf{true}\big]\\
&\text{with }M\big(\xabs{Comp.op}) \equiv\big(\xabs{a} >  0 \wedge \xabs{b} >  0, \xabs{result} >  0\big)
\end{align*}
\end{example}


\paragraph{Rules.}
The proof system is a sequent validity calculus. 
A formula is valid if it holds in every model, i.e., every state.
Let $\Gamma,\Delta$ be two finite sets of BPL formula. 
A sequent $\Gamma \Rightarrow \Delta$ represents the formula $\bigwedge\Gamma \rightarrow \bigvee\Delta$.
Fig.~\ref{fig:bplrules} gives selected rules.

Rule \textbf{asg} turns an assignment into an update. Rule \textbf{get} uses the $\mathsf{val}$ function to access the value stored in a future when turning an assignment into an update.
Rule \textbf{ret} evaluates both postconditions after termination and sets the special \abs{result} value to the return value.
Rule \textbf{skip} evaluates only the statement postcondition. Note that this only matches if no statement follows \abs{skip}.
Rule \textbf{loop} is a standard loop invariant rule, where the invariant $L$ must be annotated by the user. A final \abs{skip} is added to the loop body.
Rule \textbf{aw} handles suspension by proving the object invariant before suspension and assuming it when continuing. Update $U_\mathcal{A}$ anonymizes the heap by removing all information about \abs{heap}.
Finally, rule \textbf{call} handles calls according to $M$. The fresh variable models the future and again $\mathsf{val}$ is used to keep track of the postcondition.

\begin{figure*}
\centering
\scalebox{0.9}{
\begin{minipage}{0.52\textwidth}
\Axiom$ \Gamma \fCenter \{U\}\{\xabs{v} := \xabs{e}\}[ \xabs{s} \halfsim^\alpha \inv,M,\phi,\psi] ,\Delta$
\LeftLabel{\textbf{asg}}
\UnaryInf$ \Gamma \fCenter \{U\}[ \xabs{v = e; s} \halfsim^\alpha \inv,M,\phi,\psi] ,\Delta$
\DisplayProof
\end{minipage}
\begin{minipage}{0.47\textwidth}
\Axiom$ \Gamma \fCenter \{U\}\{\xabs{v} := \mathsf{val}(\xabs{e})\}[\xabs{s} \halfsim^\alpha \inv,M,\phi,\psi] ,\Delta$
\LeftLabel{\textbf{get}}
\UnaryInf$ \Gamma \fCenter \{U\}[ \xabs{v = e.get; s} \halfsim^\alpha \inv,M,\phi,\psi] ,\Delta$
\DisplayProof
\end{minipage}}

\vspace{2mm}
\begin{center}
{
\Axiom$ \Gamma \fCenter \{U\}\{\xabs{result} := \xabs{e}\}(\phi\wedge\psi),\Delta$
\LeftLabel{\textbf{ret}}
\UnaryInf$ \Gamma \fCenter [ \xabs{return e} \halfsim^\alpha \inv,M,\phi,\psi] ,\Delta$
\DisplayProof
}
{
\Axiom$ \Gamma \fCenter \{U\}\psi,\Delta$
\LeftLabel{\textbf{skip}}
\UnaryInf$ \Gamma \fCenter \{U\}[ \xabs{skip} \halfsim^\alpha \inv,M,\phi,\psi] ,\Delta$
\DisplayProof
}
\end{center}

\scalebox{0.85}{
\begin{minipage}{\textwidth}
\begin{prooftree}
\Axiom$ L\wedge\neg\xabs{e} \fCenter [\xabs{s'} \halfsim^\alpha \inv,M,\phi,\psi]\qquad\Gamma,\{U\}\xabs{e} \Rightarrow \{U\}L,\Delta \qquad L\wedge\xabs{e} \Rightarrow [ \xabs{s; skip} \halfsim^\alpha \inv,M,\phi,L]$
\LeftLabel{\textbf{loop}}
\UnaryInf$ \Gamma \fCenter \{U\}[ \xabs{while(e)}\{\xabs{s}\}\xabs{; s'} \halfsim^\alpha \inv,M,\phi,\psi] ,\Delta$
\end{prooftree}
\end{minipage}}
\begin{prooftree}
\Axiom$ \Gamma \fCenter \{U\}\inv,\Delta \qquad \Gamma,\{U_\mathcal{A}\}\inv \Rightarrow \{U_\mathcal{A}\}[ \xabs{s} \halfsim^\alpha \inv,M,\phi,\psi],\Delta$
\LeftLabel{\textbf{aw}}
\UnaryInf$ \Gamma \fCenter \{U\} [ \xabs{await g; s} \halfsim^\alpha \inv,M,\phi,\psi] ,\Delta$
\end{prooftree}
\scalebox{0.85}{
\begin{minipage}{\textwidth}
\begin{prooftree}
\Axiom$\Gamma \fCenter \{U\}\{\xabs{p} := \xabs{e'}\}M^{\mathsf{pr}}(\xabs{m})$
\noLine
\UnaryInf$ \Gamma,\{U\}\{\xabs{result} := \mathsf{val}(v)\}M^{\mathsf{po}}(\xabs{m}) \fCenter \{U\}\{\xabs{result} := \mathsf{val}(v)\}[\xabs{s} \halfsim^\alpha \inv,M,\phi,\psi] ,\Delta$
\LeftLabel{\textbf{call}}
\RightLabel{$v$ fresh}
\UnaryInf$ \Gamma \fCenter \{U\}[ \xabs{v = e!m(e'); s} \halfsim^\alpha \inv,M,\phi,\psi] ,\Delta$
\end{prooftree}
\end{minipage}}

\caption{Selected rules for contracts in BPL.}
\label{fig:bplrules}
\end{figure*}

The above rules are shown to be sound in~\cite{soa} and are implemented in \texttt{Crowbar}.
The code in Fig.~\ref{fig:absintro} adheres to its specification.

%% file: cabs.tex
In order to extract an ABS model annotated with appropriate specifications from a (specified) C program, we extend the approach from~\cite{WasserHH19} 
(which extracts a non-deterministic Active Objects model from C code containing underspecified behavior) 
by automatically generating some specifications which are sound by construction and generating all other specifications by translation of the specifications in the underlying C program.
In order to translate ACSL function contracts into method contracts it was also required to slightly change the manner in which function parameters were modeled, from parameters of the class to parameters of the \abs{call} method within the class.
Otherwise, simple functional properties would have required reasoning about heap properties. 
\subsubsection*{ACSL}
The \emph{ANSI/ISO C Specification Language (ACSL)}~\cite{acsl}
is a behavioral specification language for C programs,
used by the state-of-the-art \emph{Frama-C}~\cite{framac} tool suite.  
ACSL can be used to specify function contracts (pre- and postconditions), data invariants over global variables and some further constructs, such as loop invariants, statement contracts (pre- and postconditions for a single statement or block of statements), assertions or ghost code.

Function contracts consist of a \abs{requires} clause for the precondition and an \abs{ensures} clause for the postcondition.
Both clauses can be simple C expressions of arithmetic type\footnote{Full ACSL allows more operators, which we ignore for now.}, with the postcondition allowed to contain {\bf\textbackslash}\abs{result} to refer to the return value.
Additionally, an \abs{assigns} clause to specify which locations may be accessed can be given. We ignore \abs{assigns} clauses for now as they are not directly relevant for underspecified semantics.

ACSL allows two types of data invariants on global variables:
\begin{enumerate*}
 \item \emph{strong global invariants}, which hold at all times; and
 \item \emph{weak global invariants}, which hold before and after each execution of a function call and can thus equivalently be added as a requires and ensures clause to all functions.
\end{enumerate*}
We therefore focus here only on strong global invariants, in particular as these cannot be easily dealt with in Frama-C.
Furthermore, we restrict strong global invariants to properties about single variables and thus exclude relational properties.%

\subsection{From C Code to ABS (\texttt{C2ABS})}

\texttt{C2ABS}~\cite{WasserHH19} is an Eclipse plugin which extracts an ABS model from a C program. 
Here we describe how this extraction takes place.
In the next subsection we describe the novel extension of this model extraction: synthesizing specification annotations for the extracted model.
Table~\ref{table:cabs-extraction}
\begin{table*}\normalsize
 \begin{tabular}{|c|c|c|}
  \hline
  \textbf{C} & \textbf{ABS} \\
  \hline \hline
  Top-level declarations & \\
\hdashline
   Global variables & Class \xabs{Global} with methods to get/set variable values \\
 \hdashline[0.5pt/5pt]
   definition of function \xabs{f} & class \xabs{C_f} with parameter \xabs{global} \\
  \hline
  \hline
  Execution of function \xabs{f} & Execution of \xabs{call} method on object of class \xabs{C_f} \\
  \hline
  \hline
  Parameters and local variables & \\
 \hdashline
   const parameter & parameter of \xabs{call} method \\
 \hdashline[0.5pt/5pt]
   non-const parameter & parameter of \xabs{call} method stored in field \\
 \hdashline[0.5pt/5pt]
   const local variable & local variable \\
 \hdashline[0.5pt/5pt]
   non-const local variable & field \\
  \hline
  \hline
  Local const read & Direct variable/parameter access \\
  \hline
  \hline
  Other (sub-)expressions & Methods awaiting parameters and: \\
 \hdashline
   global read/write & synchronous call to \xabs{global} object (write is side effect) \\
 \hdashline[0.5pt/5pt]
   local non-const read/write & get/set value of field \\
 \hdashline[0.5pt/5pt]
   C built-in operators $\oplus$ & return result of performing $\oplus$ \\
 \hdashline[0.5pt/5pt]
   invocation of function \xabs{f} & await side effects, create \xabs{C_f} object, 
   make synchronous call to method \xabs{call} of object \\
  \hline
  \hline
  Unspecified evaluation order & Asynchronous method calls to \xabs{this} object \\
  \hline
  \hline
  Sequence points & \xabs{await} statements \\
  \hline
 \end{tabular}
 \vspace{2mm}
 \caption{Translation of C concepts into ABS}
 \label{table:cabs-extraction}
\end{table*}
details how C concepts are translated into ABS.
The basic idea is to have one Active Object which models access to global variables and further model each executed function call as its own Active Object. Within these function call objects each (sub)expression being evaluated is modeled as an asynchronous method call to itself with \abs{await} statements modeling \emph{sequence points}:
the point between evaluation of all arguments and side effects of a function call, and the call itself; the semicolon at the end of an expression statement; etc.
Access to global variables is modeled by methods making blocking calls to the \abs{global} object, while (potentially recursive) function calls are modeled by creating new Active Objects for the appropriate function and making blocking calls to these new objects.

\begin{example}\label{ex:cabs}
 Consider the function \abs{main} in Fig.~\ref{fig:cpure} and the statement \abs{return x + id_set_x(1);} inside, where there is a sequence point between evaluation of the expression and returning from the function.
 The ABS class extracted is shown in Fig.~\ref{fig:abs-main-model},
\begin{figure}[b!th]
\begin{abscode}[basicstyle=\fontsize{9}{10}\ttfamily,keywordstyle=\color{keywordcolor}\bfseries\sffamily]
class C_main(Global global) 
      implements I_main {
 Int call() {
  ...
  Fut<Int> fut_x = this!get_global_x();#\label{cabs:unspec-order-start}#
  Fut<Int> fut_set = 
    this!call_id_set_x_val_0(1);
  Fut<Int> fut_add = 
    this!op_plus_fut_fut(fut_x, fut_set);#\label{cabs:unspec-order-end}#
  await fut_x? & fut_set? & fut_add;#\label{cabs:unspec-order-await}#
  return fut_add.get;
 }
 Int get_global_x() { 
  Fut<Int> f = global!get_x(); 
  return f.get; 
 }
 Int call_id_set_x_val_0(Int arg1) {
  I_id_set_x o = new C_id_set_x(global);
  Fut<Int> f = o!call(arg1); return f.get;
 }
 Int op_plus_fut_fut(Fut<Int> fut_arg1, 
                     Fut<Int> fut_arg2) {
  await fut_arg1? & fut_arg2?;
  Int arg1 = fut_arg1.get; 
  Int arg2 = fut_arg2.get;
  return arg1 + arg2;
 }
}\end{abscode}
\caption{Class \abs{C_main} extracted from function \abs{main} in Fig.~\ref{fig:cpure}}
\label{fig:abs-main-model}
\end{figure}
where the method \abs{call} models function execution and lines~\ref{cabs:unspec-order-start}-\ref{cabs:unspec-order-end} model the unspecified evaluation order of the the expression \abs{x + id_set_x(1)} with the \abs{await} at line~\ref{cabs:unspec-order-await} allowing non-deterministic choice in which order the methods to \abs{this} are executed in.
Once all futures have been resolved, the \abs{await} regains control, modeling the sequence point before returning. The method \abs{call} then returns the value of the addition.
The method \abs{get_global_x} models the memory access, by making a synchronous call\footnote{An asynchronous call to an object in another object immediately followed by a \abs{get}.} to the \abs{global} parameter of the class, requesting the value of \abs{x}.
The method \abs{call_id_set_x_val_0} models a call to the function \abs{id_set_x} with an argument evaluated at compile time and zero side effects from evaluating its argument\footnote{If the argument were a future or side effects (modeled as futures) were present, the method would immediately await resolution of all these futures.}.
This is done by first creating a new \abs{C_id_set_x} object with access to the same \abs{global} object and then making a synchronous call to the \abs{call} method of that object with the evaluated function arguments as parameters.
Finally, the method \abs{op_plus_fut_fut} models the addition of two subexpressions evaluated at runtime and therefore modeled as futures. First, the method awaits the resolution of its subexpressions, then returns the sum.
While the three methods can be executed in arbitrary (and interleaving) order, the only visible difference depends on the order of \abs{get_global_x} and \abs{call_id_set_x_val_0}, as \abs{op_plus_fut_fut} immediately awaits resolution of the other two methods.
\end{example}

\subsection{Automatically Synthesizing Specifications}

Due to the automated nature in which function-modelling classes and helper methods are generated, we can synthesize some specifications directly.
For others we require ACSL specification of the underlying C program.

\subsubsection{Auto-generate specifications related to global object}

%
As each function-modelling class receives the \xabs{global} object as a parameter, uses it to access global variables and passes it on when instantiating any further function-modelling classes, we must (at least) specify that this class parameter (and field) is never \xabs{null}.
To this end all function-modelling classes are specified with:

\noindent
\begin{abscode}[numbers=none]
[Spec : Requires(global != null)]
[Spec : ObjInv(global != null)]
\end{abscode}

\subsubsection{Auto-generate precise postconditions for operator methods}

\texttt{C2ABS}-generated methods from C built-in operators $\oplus$ all perform the same basic steps:
await resolution of all future parameters and then return the result of performing $\oplus$ on the (resolved) parameters.
Precise postcondition specifications for each of these methods can therefore be generated automatically, by ensuring that the result of the method is equal to the result of performing $\oplus$ on the (resolved) parameters.
All C operator method declarations in interfaces are thus automatically annotated with appropriate postcondition specifications.

\begin{example}
 The interface \abs{I_main} in the model extracted from the program in Fig.~\ref{fig:cpure} contains the following annotated method declaration:
 
\noindent
\begin{abscode}[numbers=none]
[Spec : Ensures(valueof(fut_arg1) + valueof(fut_arg2) == result)]
Int op_plus_fut_fut(Fut<Int> fut_arg1, 
                    Fut<Int> fut_arg2);
\end{abscode}
\end{example}

\subsubsection{Translate ACSL requires/ensures function contracts}

ACSL requires/ensures clauses specify (relational) restrictions upon the function arguments and functional guarantees for the result.
Following similar steps to those for extracting C expressions---simplified somewhat due to lack of side effects---these can be converted
into pre- and postconditions of the \abs{call} method in the interface modelling the function.
Additionally, similar pre- and postconditions are added to the indirect call methods in any interfaces modelling functions calling the specified function.
When an argument to an indirect call is a future value, the pre- and postconditions must be formulated to hold for the resolved argument.

\begin{example}
 Given the specified function \abs{id_set_x} at line \ref{id_set_x_spec} in Fig.~\ref{fig:example1}:
 
\noindent
\begin{ccode}[firstnumber=2]
int id_set_x(int val) 
/*@ requires val == 1; ensures \result == 1; @*/ {
\end{ccode}

We annotate both the \abs{call} method in \abs{I_id_set_x} and the \abs{call_id_set_x_val} method in \abs{I_main} with the following specifications:

\noindent
\begin{abscode}[numbers=none]
[Spec : Requires(val == 1)]
[Spec : Ensures(result == 1)]
\end{abscode}
\end{example}

\subsubsection{Translate ACSL strong global invariants}


While a strong global invariant must hold at every point in the program, it suffices to \emph{check} that it holds at program start and whenever the global variable is changed.
The ACSL invariant is translated as above and added as an object invariant in the \abs{Global} class and as preconditions on the argument of all setter methods for said variable.
When the argument to indirect setters outside of \abs{Global} is a future value, the precondition must be formulated to hold for the resolved argument.
In order to \emph{use} the invariant, we add postconditions to all getter methods for the variable.

\begin{example}
 Given the strong global invariant at line \ref{strong_global_invariant} in Fig.~\ref{fig:example1} that \abs{x == 0 || x == 1}, the global state is modeled as
 the code in Fig.~\ref{fig:stronginv}.
 Additionally, \abs{I_id_set_x} and \abs{I_main} contain the annotated method declarations in the lower code in Fig.~\ref{fig:stronginv}.
 
 \begin{figure}
\begin{abscode}[numbers=none,basicstyle=\fontsize{9}{10}\ttfamily,keywordstyle=\color{keywordcolor}\bfseries\sffamily]
interface Global {
 [Spec : Ensures(result == 0||result==1)] 
 Int get_x();
 [Spec : Requires(arg == 0||arg == 1)] 
 Unit set_x(Int arg);
}
[Spec : ObjInv(this.x == 0 || this.x == 1)]
class Global implements Global {
 Int x = 0;
 Int get_x() { return this.x; }
 Unit set_x(Int arg) { 
  this.x = arg; 
  return unit; 
 }
}
\end{abscode}
\begin{abscode}[numbers=none,basicstyle=\fontsize{9}{10}\ttfamily,keywordstyle=\color{keywordcolor}\bfseries\sffamily]
[Spec:Requires(arg == 0 || arg == 1)]
Unit set_global_x_val(Int arg);
[Spec:Requires(valueof(fut_arg) == 0||valueof(fut_arg) == 1)]
Unit set_global_x_fut(Fut<Int> fut_arg);
[Spec:Ensures(result == 0||result == 1)] 
Int get_global_x();
\end{abscode}
 \caption{Example for translating strong global invariants.}
 \label{fig:stronginv}
 \end{figure}
\end{example}

\subsubsection{Use ABS functions in lieu of ACSL logic functions}

ACSL allows pure \emph{logic functions} to be defined (inductively or axiomatically) and called in ACSL specifications.
Translating these definitions is outside of the scope of this work and we therefore instead allow ABS functions to be called directly in ACSL specifications.
If the ABS function is not inside the standard library, it must be defined inside an ACSL-style comment in the C program.

\subsubsection{Scope}
The C Standard lists 52 cases of unspecified behavior~\cite[Annex. J.1]{ISO:C99}. However, most of these cases are not relevant to functional verification of runtime semantics, e.g., unspecified behavior of macros; or concern well-investigated elements outside of the considered language fragment, such as floating points and string literals; or concern deprecated features of old compilers for rare hardware, such as the use of negative zeros in integer types.
Our focus is therefore on those cases that touch on core aspects of the runtime semantics and are relevant for almost all programs: order of subexpression and side effect evaluation (except for some operators such as $\&\&$)~\cite[6.5]{ISO:C99}, of function argument evaluation~\cite[6.5.2.2]{ISO:C99} and of evaluation of complex assignments~\cite[6.5.16]{ISO:C99}.
All these aspects can be handled by our approach and reduced to non-determinism of concurrent systems.

%% file: crowbar.tex
\texttt{Crowbar} is a verification system that implements symbolic execution (SE) 
i.e., the step-wise execution of statements to generate a set of first-order logic formulas. 
Validity of all generated formulas implies safety of the method.
The resulting formulas are output in SMT-LIB~\cite{Barrett2010TheSS} format and passed to solvers such as Z3. 

Contrary to its predecessor, KeY-ABS~\cite{DinBH15}, \texttt{Crowbar} performs no rewriting during SE, but normalizes ABS input models after parsing.
E.g., all \abs{get} statements assign to variables after normalization. 
Normalization and the use of the builtin theories of Z3 drastically reduces the number of rules for SE: \texttt{Crowbar} has 11 rules for SE and requires no BPL-specific first-order theory. KeY-ABS uses 24 SE rules, covers a smaller fragment of syntax than \texttt{Crowbar}, and has over 30 rules that implement a theory of event traces.

\paragraph{Deadlock Checking.}
\texttt{Crowbar} implements a lightweight deadlock checker for ABS that contrary to existing deadlock checkers for ABS~\cite{avocs}, requires no main block:
The structural deadlock analysis deduces which methods cannot be part of a deadlock \emph{for any program}:
A deadlock is a cycle of dependencies caused by future (and condition) synchronizations~\cite{avocs} and is analyzed in terms of cycles in dependency graphs between synchronizations, objects and methods. 
Any method that contains no synchronization cannot be part of any dependency cycle, it is \emph{structurally deadlock-free}. 
Similarly, all methods that only call deadlock-free methods and synchronize only on their futures are not part of any deadlock.

\begin{example}
Consider Ex.~\ref{ex:lang}. 
If the implementation of \mbox{\abs{CompC.op}} contains no blocks or call, e.g., the statement \mbox{\abs{return a*b}}, 
then we can show deadlock freedom.

\abs{CompC.op} is structurally deadlock-free:
it contains no synchronization or suspension.
\abs{C.fold} depends only on \mbox{\abs{CompC.op}} and is thus not part of any deadlock.
\end{example}

%% file: fib.tex
\FloatBarrier
Underspecified behavior lurks at every binary operation\footnote{Underspecified behavior also lurks at many function calls.} and can have subtle effects in larger systems.
To evaluate our verification approach, we use an extreme case of underspecification,
investigating the C program%
\footnote{Adapted from an idea on Derek Jones's \emph{The Shape of Code} blog at:\\ \mbox{\url{shape-of-code.coding-guidelines.com/2011/06/18/fibonacci-and-jit-compilers/}}}
in Fig.~\ref{fig:fib-c}
\begin{figure}[t]
\noindent
\begin{ccode}
//@ ABS def Int fib(Int n) = if n <= 2 then 1 
//@                          else fib(n-1) + fib(n-2);#\label{one-to-fib:fib-def}#

/*@ strong global invariant x == 0 || x == 1; @*/ int x;#\label{one-to-fib:x-decl}#

//@ ensures \result == val;
int id_set_x(const int val) 
{ x=1; return val; }
//@ ensures \result == 1 || \result == 2;
int one_or_two(void) { #\label{one-to-fib:plus-1}#
    x=0; 
    return x + id_set_x(1); 
}
//@ ensures \result == val - 1 || \result == val;
int pred_or_id(const int val) { #\label{one-to-fib:plus-minus}#
    x=0; 
    return val - x + id_set_x(0); 
}
//@ ensures \result >= 1 && \result <= fib(n);
int one_to_fib(const int n) {
 if (n > 3) 
  return one_to_fib(n-2) 
        + pred_or_id(one_to_fib(n-1));#\label{one-to-fib:plus-2}#
 else if (n == 3) return one_or_two(); 
 else return 1; }
\end{ccode}
\caption{Calculate a number between 1 and the \abs{n}th Fibonacci number in C}
\label{fig:fib-c}
\end{figure}
containing a function whose result heavily depends on unspecified evaluation order.
The function in question is declared as \mbox{\abs{int one_to_fib(int n)}} and should calculate a number between 1 and the nth Fibonacci number.
The base cases are for inputs 1 and 2 (as well as all non-positive inputs), which return 1; as well as for input 3, which returns either 1 or 2 in the same manner as the program in Figure~\ref{fig:cpure}.
Otherwise, \mbox{\abs{one_to_fib(n)}} returns the sum of \abs{one_to_fib(n-2)} and \abs{one_to_fib(n-1)} with a potential decrement of 1 in the function \abs{pred_or_id} ensuring that 1 is always a potential result, as otherwise $\{1,\ldots,Fib(n-1)\} + \{1,\ldots,Fib(n-2)\}$ $=$ $\{2,\ldots,Fib(n)\}$. 

Verification of this program is a challenging task due to the extensive non-determinism.
In~\cite{WasserHH19} the extracted model for this program was exhaustively checked for inputs up to 5, validating that all possible outputs (and no outputs outside this range) could be produced.
Later experiments with an enhanced model extraction process partially validated models for inputs up to 10.
In this work we verify that no outputs outside of the range are produced for any (valid) inputs.%
\footnote{The semantics of the program are underspecified but \emph{not} undefined.} 
The annotated extracted model for this C program can be found in the appendix. 
The ABS function definition inside the ACSL-style specification in line~\ref{one-to-fib:fib-def} is copied verbatim into the model,
the helper methods for \abs{+} (used in lines~\ref{one-to-fib:plus-1}, \ref{one-to-fib:plus-minus} and \ref{one-to-fib:plus-2}) and \abs{-} (line~\ref{one-to-fib:plus-minus}) receive precise specifications,
the strong global invariant on \abs{x} at line~\ref{one-to-fib:x-decl} produces specifications throughout the model (\abs{Global} interface and class, plus indirect getter and setter methods of other interfaces),
while the \abs{call} methods and their indirect callers are specified with translations of the contracts for the matching functions.
As the program does not contain a \abs{main} method and is not executable, so the model it produces is therefore also not executable: the main block in the extracted model is empty.
As we are focused on proving a property of \abs{one_to_fib} in general, rather than for a specific actual call, this non-executability is not a problem.
This shows an additional strength of our approach, in that we can analyze \emph{library} calls in isolation, rather than only being able to analyze a complete program.

\texttt{Crowbar} can close all proof obligations of the extracted model \emph{automatically}.
Note that we prove the following \emph{for all inputs} to \abs{one_to_fib}.
\begin{theorem}\label{lem1}
The extracted model is safe with respect to its specification.
\end{theorem}

\paragraph*{Deadlock Freedom.}
Running \texttt{Crowbar} performs a simple analysis for structurally deadlock-free methods and returns all methods for which it cannot deduce it.

For the extracted model it returns 9 such methods. 
These are the methods that take futures as parameters, which is not supported by the deadlock analysis in \texttt{Crowbar}, and methods depending on these methods.
However, all futures that are passed as parameters are always futures of free methods. Thus we can state the following lemma, which is proven in the appendix.
\begin{lemma}\label{lem2}
The extracted model is deadlock free for every extractable main block. 
\end{lemma}

%% file: conclusion.tex
We have demonstrated a novel approach combining model extraction with deductive verification of a distributed active objects model in order to verify C programs with underspecified behavior by reducing the non-determinism of underspecification to non-determinism of parallelism.
We have extended the \texttt{C2ABS} tool---which already gives C a formal semantics in terms of Active Objects---
to automatically translate a large subset of ACSL specifications into BPL specifications and implemented the \texttt{Crowbar} tool based on~\cite{Kamburjan19} in order to verify the specified model and analyze it for deadlock freedom.
Using a complex case study that exemplifies the challenges for verification of underspecified programs we showed that our approach of model extraction and verification is \emph{fully automatic}.
We reused a standard logic and deadlock analysis for ABS and did not need special amendments for underspecified behavior after the extraction. 

This is the first implementation and application of BPL. 
It was expected~\cite{thesisek} that the encoding of contracts and invariants in the proof system is more suited for automatization than using a special variable to keep track of the history.
We take the fully automatic proofs, which were not achieved with the history-based approach for far simpler programs, as an indication that this is indeed the case.

\paragraph*{Future Work.}
For formalized parallelization of C code, we plan to integrate a formal, logic-based dependences analysis~\cite{BubelHT19} and to consider further cases of underspecification of a larger fragment of C, e.g., in list initializers.
The newest version of \texttt{C2ABS} uses different model extraction strategies~\cite{WasserHH21} and we will investigate using \texttt{Crowbar} to verify these models as well.
In cases where the input C program is not completely specified, we envisage generating the missing object invariants and method contracts automatically via counter-example guided refinement techniques~\cite{CEGAR} using the failed \texttt{Crowbar} proofs.



%% file: appendix.tex
\subsection{Overview Local Layer}

\begin{definition}
A \emph{state} $\sigma$ is a mapping from fields and variables to values.
An \emph{event} $\mathsf{ev}$ denotes visible side-effects and is defined as:
\begin{align*}
\mathsf{ev} ::=& \invocev(X,X',f,\methodname,\many{\xabs{e}}) \sep \invocrev(X,f,\methodname,\many{\xabs{e}}) \sep \suspev(X,f,\many{f})\\
\sep& \susprev(X,f,\many{f}) \sep \resolvev(X,f,\xabs{e}) \sep \resolvrev(X,f,f',\xabs{e}) \sep \marker\sep \noev
\end{align*}
A \emph{local trace} $\theta = \mathsf{sc} \triangleright \mathsf{hs}$ is a pair of a set of expressions $\mathsf{sc}$ and a non-empty sequence $\mathsf{hs}$ that alternates between states and events. 
\end{definition}

Events are used to expose effects to the object and system during composition. Here, we only use events for method calls and future access. 
The event \invocev denotes a method call of object $X$ on object $X'$ with method $\methodname$, future $f$ and parameters $\many{\xabs{e}}$.
The event \invocrev denotes the corresponding start of the process.
Events \suspev and \susprev denote suspension and reactivation of object $X$ computing future $f$ and synchronizing on the futures $\many{f}$.
The event \resolvev denotes that object $X$ resolves future $f$ with return value $\xabs{e}$ and \resolvrev the read of object $X$ (computing future $f$) from future $f'$, which was resolved with value \abs{e}.
The event \marker is a special symbol that is a placeholder for other processes to run. It is needed for technical reasons by the global layer of the semantics.
Finally, \noev denotes a step with no visible communication.

The \emph{selection condition} $\mathsf{sc}$ of a trace is a set of boolean expressions. 
A trace is only considered for global composition if all expressions in its selection condition evaluate to true at the moment of selection.
The \emph{history} $\mathsf{hs}$ of a trace describes the state changes and side-effects of a statement.

\begin{definition}
The \emph{semantics} of statements \abs{s} are defined by a function $\eval{\xabs{s}}{\sigma}$ that maps each statement and state to a set of possible local traces.
Fig.~\ref{fig:semtrace} shows the most important rules for statements. Evaluation of expressions is written $\eval{\xabs{e}}{\sigma}$. 
\end{definition}

The semantics map each statement to a set of possible traces. During execution, the global layer gradually selects one of the traces of the method body for execution.
The semantics of an assignment is a trace without selection condition and a history where the given state is updated according to the assignment. No visible communication\footnote{In this work, we do not consider field accesses as visible communication.} happens. Analogously for termination.
The semantics of a future read are infinitely many traces, one for each possible read value. Similarly for method calls. The global semantics ensures that only the correct one of these traces is chosen.
The semantics of a synchronization emits three events: loss of control (\awaitev), a placeholder for other traces (\marker) and regaining control (\awaitrev).
Only the fields are changed after reactivation, the local variables stay the same. 
Sequential composition ensures that the second subtrace is executed from the last state of the first subtrace. The first state of the second subtrace is removed, because it is identical to the last state of the first subtrace.
Branching and repetition are straightforward. For the convenience of the reviewers, Sec.~\ref{appendix:examples} gives an example of the semantics.
\begin{figure}[b!th]
\scalebox{0.9}{
    \begin{minipage}{\textwidth}
\begin{align*}
\eval{\xabs{v = e}}{\sigma} =& \big\{\emptyset \triangleright \bigsequence{\sigma,\noev,\sigma[\xabs{v} \setminus \xabs{e}]}\big\}
\quad\eval{\xabs{return e}}{\sigma} = \big\{\emptyset \triangleright \sequence{\sigma,\resolvev(X,f,\eval{\xabs{e}}{\sigma}),\sigma})\big\}\\
\eval{\xabs{v = e.get}}{\sigma} =& \{\emptyset \triangleright \sequence{\sigma,\resolvrev(X,f,\eval{e}{\sigma},v),\sigma[\xabs{v} \setminus v] \sep v \in \mathsf{T}(\xabs{v})}\}\\
\eval{\xabs{v = e!m(}\many{\xabs{e}}\xabs{)}}{\sigma} =& \Big\{\emptyset \triangleright \sequence{\sigma,\invocev(X,\eval{e}{\sigma},f',\xabs{m},\many{\eval{\xabs{e}}{\sigma}}),\sigma[\xabs{v} \setminus f']} \sep f' \text{fresh}\Big\}\\
\eval{\xabs{await}~\bigwedge_i\xabs{e}_i}{\sigma} =& \Big\{\emptyset \triangleright\bigsequence{\sigma,\awaitev(X,f,\{\eval{\xabs{e}_i}{\sigma})\},\sigma,\marker,\sigma',\awaitrev(X,f,\{\eval{\xabs{e}_i}{\sigma}\}),\sigma'}\\\vspace{-4mm}
&\hspace{2mm}\sep\text{for each field \xabs{f}: }\sigma'(\xabs{f}) = v_\xabs{f}\text{ for some $v_\xabs{f} \in \mathsf{T}(\xabs{f})$}\\
&\hspace{5mm}\text{for each variable \xabs{v}}: \sigma'(\xabs{v}) = \sigma(\xabs{v})\Big\}\\
\eval{\xabs{s}_1\xabs{;s}_2}{\sigma} =&\Big\{\big(\mathsf{sc}_1\cup\mathsf{sc}_2\big) \triangleright \big(\mathsf{hs}_1\circ\mathsf{tail}(\mathsf{hs}_2)\big) \sep \\
&\big(\mathsf{sc}_1\triangleright\mathsf{hs}_1\big) \in \eval{\xabs{s}_1}{\sigma}, \big(\mathsf{sc}_2\triangleright\mathsf{hs}_2\big) \in \eval{\xabs{s}_2}{\mathsf{last}(\mathsf{hs}_1)}\Big\}\\ 
\eval{\xabs{if(e)}\{\xabs{s}_1\}\xabs{else}\{\xabs{s}_2\}}{\sigma} =&\big\{\mathsf{sc} \cup \{\neg\xabs{e}\} \triangleright \mathsf{hs} \sep \mathsf{sc}\triangleright\mathsf{hs} \in \eval{\xabs{s}_1}{\sigma}\big\}
\big\{\mathsf{sc} \cup \{\xabs{e}\} \triangleright \mathsf{hs} \sep \mathsf{sc}\triangleright\mathsf{hs} \in \eval{\xabs{s}_2}{\sigma}\big\}\\
\eval{\xabs{while(e)}\{\xabs{s}\}}{\sigma} =& \eval{\xabs{if(e)}\{\xabs{s; while(e)}\{\xabs{s}\}\}\xabs{else}\{\xabs{skip}\}}{\sigma}
\end{align*}
\end{minipage}}
\caption{Selected semantics rules. $\mathsf{tail}$ drops the first element of a history. Object $X$ and future $f$ are implicitly known. $\mathsf{T}(\cdot)$ is the data type of a field or variable.}
\label{fig:semtrace}
\end{figure}

\subsection{Contracts as Trace Properties}
The trace logic contains no modalities; instead it has formulas $[i] \vdash \phi$ to express that the $i$th element of the trace is a state where the BPL formula $\phi$ holds
and formulas $[i] \doteq \mathsf{evt}$ to express that the $i$th element of the trace is a is an event of form $\mathsf{evt}$.
It allows to quantify over positions in the trace with the sort $\mathbf{I}$.
We use $\_$ as a shortform for a logical variable that occurs nowhere else, with the smallest possible existantial quantifier. 
For a formal treatment we refer to~\cite{soa}.
\begin{definition}[Semantics of Contracts]
Let $D$ be the return type of \abs{C.m} and $a(\xabs{m'})$ the number of all parameters of \abs{m'} and $D_i(\xabs{m'})$ the type of the $i$th parameters.
The semantics $\alpha(I,M,\phi,\psi)$ of a contract for \abs{C.m} is defined as follows. 
\scalebox{0.875}{
    \begin{minipage}{\textwidth}
\begin{align*}
\forall i \in \mathbf{I}.~
\Big(&(\big([i] \doteq \awaitev(\_,\_,\_) \vee [i] \doteq \awaitev(\_,\_,\_) \rightarrow [i+1] \vdash I\big)\\
\wedge&\big(\exists t\in D.~[i] \doteq \resolvev(\_,\_,t) \rightarrow [i+1] \vdash \phi[\xabs{result} \setminus t]\big) \wedge\big(\forall j \in \mathbf{I}.~j \leq i \rightarrow [i] \vdash \psi\big)\\
\wedge&\big(\bigwedge_{\xabs{m'} \in \mathbf{dom}(M)}\exists t_1\in D_1(\xabs{m'}).~\cdots\exists t_{a(\xabs{m'})}\in D_{a(\xabs{m'})}(\xabs{m'}).~\\
&\hspace{20mm}\big([i] \doteq \invocev(\_,\_,\_,\xabs{m'},\sequence{t_1,\dots,t_{a(\xabs{m'})}}) \rightarrow [i+1]\vdash \widetilde{M^{\mathsf{pr}}(\xabs{m'})}_{t_1,\dots,t_{a(\xabs{m'})}}\big)\big)\Big)
\end{align*}
\end{minipage}}
\end{definition}
The semantics formalize the intuition described above. The first line describes that before and after suspension the object invariant $I$ holds.
The second line formalizes method (per \resolvev) and statement (per last element) postconditions. The last two lines formalize call conditions, where 
$\widetilde{M^{\mathsf{pr}}(\xabs{m'})}_{t_1,\dots,t_{a(\xabs{m'})}}$ replaces the abstract parameters of $M^{\mathsf{pr}}(\xabs{m'})$ with the logical variables $t_1,\dots,t_{a(\xabs{m'})}$.

%% file: appendix_ex.tex
The following examples illustrates the trace-based semantics of ABS:
\begin{example}
Consider the state $\sigma = \{\xabs{this.f} \mapsto 0,\xabs{x} \mapsto \mathit{fut}\}$, where \abs{this.f} is a field, \abs{x}
a local variable and $\mathit{fut}$ some future.
The semantics of the statement \abs{await x?; this.f = this.f*this.f; return this.f;} from $\sigma$ is as follows
\begin{align*}
&\eval{\xabs{await x?; this.f = this.f*this.f; return this.f;}}{\sigma} = \\
\Big\{\phantom{\sep}&\emptyset \triangleright \big\langle\sigma,\suspev(X,f,\mathit{fut}),\sigma,\marker,\sigma,\susprev(X,f,\mathit{fut}),\\
&\hspace{7mm}\sigma',\noev,\sigma'',\resolvev(X,f,v*v),\sigma''\big\rangle \\
\sep& \sigma' = \sigma[\xabs{this.f} \mapsto v], \sigma'' = \sigma[\xabs{this.f} \mapsto v*v], v \in \mathbb{Z} \Big\}
\end{align*}
The set is infinite, yet in each run the global layer of the semantics selects only one --- the one where \abs{this.f} is set correctly, i.e., according to the current state. 
\end{example}

%% file: fib-abs.tex
\noindent
\begin{abscode}
module TestModule;

data Spec =
  ObjInv(Bool) | Ensures(Bool) | Requires(Bool) | WhileInv(Bool);

def Int fib(Int n) = 
  if ( n <= 2 ) then 1 else ( fib(( n - 1 )) + fib(( n - 2 )) );

interface Global {
  [Spec : Ensures(( ( result == 0 ) || ( result == 1 ) ))]
  Int get_x();
  [Spec : Requires(( ( value == 0 ) || ( value == 1 ) ))]
  Unit set_x(Int value);
}
[Spec : ObjInv(( ( this.x == 0 ) || ( this.x == 1 ) ))]
class Global implements Global {
  Int x = 0;
  Int get_x(){ return this.x; }
  Unit set_x(Int value){ this.x = value; }
}

interface I_id_set_x {
  [Spec : Ensures(( result == val ))]
  Int call(Int val);
  [Spec : Requires(( ( value == 0 ) || ( value == 1 ) ))]
  Unit set_global_x_val(Int value);
}
[Spec : Requires(( this.global != null ))]
[Spec : ObjInv(( this.global != null ))]
class C_id_set_x(Global global) implements I_id_set_x {
  Int call(Int val){
    Bool returnFlag = False;
    Int funcResult = 0;
    {
      Fut<Unit> tmp_2 = this!set_global_x_val(1);
      await tmp_2?;
      funcResult = val;
      returnFlag = True;
    }
    return funcResult;
  }
  Unit set_global_x_val(Int value){
    Fut<Unit> futureResult = this.global!set_x(value);
    futureResult.get;
  }
}
\end{abscode}

\noindent
\begin{abscode}[firstnumber=last]
interface I_one_or_two {
  [Spec : Ensures(( ( result == 1 ) || ( result == 2 ) ))]
  Int call();
  [Spec : Requires(( ( value == 0 ) || ( value == 1 ) ))]
  Unit set_global_x_val(Int value);
  [Spec : Ensures(( result == ( valueOf(fut_arg1) + valueOf(fut_arg2) ) ))]
  Int op_plus_fut_fut(Fut<Int> fut_arg1, Fut<Int> fut_arg2);
  [Spec : Ensures(( ( result == 0 ) || ( result == 1 ) ))]
  Int get_global_x();
  [Spec : Ensures(( result == arg1 ))]
  Int call_id_set_x_val_0(Int arg1);
}
[Spec : Requires(( this.global != null ))]
[Spec : ObjInv(( this.global != null ))]
class C_one_or_two(Global global) implements I_one_or_two {
  Int call(){
    Bool returnFlag = False;
    Int funcResult = 0;
    {
      Fut<Unit> tmp_4 = this!set_global_x_val(0);
      await tmp_4?;
      Fut<Int> tmp_5 = this!get_global_x();
      Fut<Int> tmp_6 = this!call_id_set_x_val_0(1);
      Fut<Int> tmp_7 = this!op_plus_fut_fut(tmp_5, tmp_6);
      await tmp_7?;
      funcResult = tmp_7.get;
      returnFlag = True;
    }
    return funcResult;
  }
  Int op_plus_fut_fut(Fut<Int> fut_arg1, Fut<Int> fut_arg2){
    await fut_arg1? & fut_arg2?;
    Int arg1 = fut_arg1.get;
    Int arg2 = fut_arg2.get;
    return ( arg1 + arg2 );
  }
  Int get_global_x(){
    Fut<Int> futureResult = this.global!get_x();
    Int funcResult = futureResult.get;
    return funcResult;
  }
  Int call_id_set_x_val_0(Int arg1){
    I_id_set_x tmp_8 = new C_id_set_x(this.global);
    Fut<Int> tmp_9 = tmp_8!call(arg1);
    Int tmp_10 = tmp_9.get;
    return tmp_10;
  }
\end{abscode}

\noindent
\begin{abscode}[firstnumber=last]
  Unit set_global_x_val(Int value){
    Fut<Unit> futureResult = this.global!set_x(value);
    futureResult.get;
  }
}

interface I_pred_or_id {
  [Spec : Ensures(( ( result == ( val - 1 ) ) || ( result == val ) ))]
  Int call(Int val);
  [Spec : Requires(( ( value == 0 ) || ( value == 1 ) ))]
  Unit set_global_x_val(Int value);
  [Spec : Ensures(( result == ( arg1 - valueOf(fut_arg2) ) ))]
  Int op_minus_val_fut(Int arg1, Fut<Int> fut_arg2);
  [Spec : Ensures(( result == arg1 ))]
  Int call_id_set_x_val_0(Int arg1);
  [Spec : Ensures(( result == ( valueOf(fut_arg1) + valueOf(fut_arg2) ) ))]
  Int op_plus_fut_fut(Fut<Int> fut_arg1, Fut<Int> fut_arg2);
  [Spec : Ensures(( ( result == 0 ) || ( result == 1 ) ))]
  Int get_global_x();
}
[Spec : Requires(( this.global != null ))]
[Spec : ObjInv(( this.global != null ))]
class C_pred_or_id(Global global) implements I_pred_or_id {
  Int call(Int val){
    Bool returnFlag = False;
    Int funcResult = 0;
    {
      Fut<Unit> tmp_12 = this!set_global_x_val(0);
      await tmp_12?;
      Fut<Int> tmp_13 = this!get_global_x();
      Fut<Int> tmp_14 = this!op_minus_val_fut(val, tmp_13);
      Fut<Int> tmp_15 = this!call_id_set_x_val_0(0);
      Fut<Int> tmp_16 = this!op_plus_fut_fut(tmp_14, tmp_15);
      await tmp_16?;
      funcResult = tmp_16.get;
      returnFlag = True;
    }
    return funcResult;
  }
  Int call_id_set_x_val_0(Int arg1){
    I_id_set_x tmp_17 = new C_id_set_x(this.global);
    Fut<Int> tmp_18 = tmp_17!call(arg1);
    Int tmp_19 = tmp_18.get;
    return tmp_19;
  }
\end{abscode}

\noindent
\begin{abscode}[firstnumber=last]
  Unit set_global_x_val(Int value){
    Fut<Unit> futureResult = this.global!set_x(value);
    futureResult.get;
  }
  Int op_minus_val_fut(Int arg1, Fut<Int> fut_arg2){
    await fut_arg2?;
    Int arg2 = fut_arg2.get;
    return ( arg1 - arg2 );
  }
  Int op_plus_fut_fut(Fut<Int> fut_arg1, Fut<Int> fut_arg2){
    await fut_arg1? & fut_arg2?;
    Int arg1 = fut_arg1.get;
    Int arg2 = fut_arg2.get;
    return ( arg1 + arg2 );
  }
  Int get_global_x(){
    Fut<Int> futureResult = this.global!get_x();
    Int funcResult = futureResult.get;
    return funcResult;
  }
}

interface I_one_to_fib {
  [Spec : Ensures(( ( result >= 1 ) && ( result <= fib(n) ) ))]
  Int call(Int n);
  [Spec : Ensures(( ( result == ( valueOf(fut_arg1) - 1 ) ) || ( result == valueOf(fut_arg1) ) ))]
  Int call_pred_or_id_fut_0(Fut<Int> fut_arg1);
  [Spec : Ensures(( ( result >= 1 ) && ( result <= fib(arg1) ) ))]
  Int call_one_to_fib_val_0(Int arg1);
  [Spec : Ensures(( result == ( valueOf(fut_arg1) + valueOf(fut_arg2) ) ))]
  Int op_plus_fut_fut(Fut<Int> fut_arg1, Fut<Int> fut_arg2);
  [Spec : Ensures(( ( result == 1 ) || ( result == 2 ) ))]
  Int call_one_or_two_0();
}
[Spec : Requires(( this.global != null ))]
[Spec : ObjInv(( this.global != null ))]
class C_one_to_fib(Global global) implements I_one_to_fib {
  Int call(Int n){
    Bool returnFlag = False;
    Int funcResult = 0;
    {
      Int tmp_20 = if ( n > 3 ) then 1 else 0;
      if ( ( tmp_20 != 0 ) ){
        Fut<Int> tmp_21 = this!call_one_to_fib_val_0(( n - 2 ));
        Fut<Int> tmp_22 = this!call_one_to_fib_val_0(( n - 1 ));
        Fut<Int> tmp_23 = this!call_pred_or_id_fut_0(tmp_22);
        Fut<Int> tmp_24 = this!op_plus_fut_fut(tmp_21, tmp_23);
\end{abscode}

\noindent
\begin{abscode}[firstnumber=last]
        await tmp_24?;
        funcResult = tmp_24.get;
        returnFlag = True;
      } else {
        Int tmp_31 = if ( n == 3 ) then 1 else 0;
        if ( ( tmp_31 != 0 ) ){
          Fut<Int> tmp_32 = this!call_one_or_two_0();
          await tmp_32?;
          funcResult = tmp_32.get;
          returnFlag = True;
        } else {
          funcResult = 1;
          returnFlag = True;
        }
      }
    }
    return funcResult;
  }
  Int call_pred_or_id_fut_0(Fut<Int> fut_arg1){
    await fut_arg1?;
    Int arg1 = fut_arg1.get;
    I_pred_or_id tmp_28 = new C_pred_or_id(this.global);
    Fut<Int> tmp_29 = tmp_28!call(arg1);
    Int tmp_30 = tmp_29.get;
    return tmp_30;
  }
  Int call_one_to_fib_val_0(Int arg1){
    I_one_to_fib tmp_25 = new C_one_to_fib(this.global);
    Fut<Int> tmp_26 = tmp_25!call(arg1);
    Int tmp_27 = tmp_26.get;
    return tmp_27;
  }
  Int op_plus_fut_fut(Fut<Int> fut_arg1, Fut<Int> fut_arg2){
    await fut_arg1? & fut_arg2?;
    Int arg1 = fut_arg1.get;
    Int arg2 = fut_arg2.get;
    return ( arg1 + arg2 );
  }
  Int call_one_or_two_0(){
    I_one_or_two tmp_33 = new C_one_or_two(this.global);
    Fut<Int> tmp_34 = tmp_33!call();
    Int tmp_35 = tmp_34.get;
    return tmp_35;
  }
}

{ /* empty main block as no main function present */ }
\end{abscode}